\documentclass[prl,amsmath,amssymb, twocolumn, showpacs]{revtex4}

\usepackage{dcolumn}
\usepackage{bm}
\usepackage{graphicx}

\begin{document}

\title{Embedding dissipation and decoherence in unitary evolution schemes}

\author{A.\ R.\  P. Rau$^{*}$ and R.\ A. Wendell}
\affiliation{Department of Physics and Astronomy, Louisiana State University,
Baton Rouge, Louisiana 70803-4001}


\begin{abstract}
Dissipation and decoherence, and the evolution from pure to mixed states in
quantum physics are handled through master equations for the density matrix.
By embedding elements of this matrix in a higher-dimensional Liouville-Bloch
equation, the methods of unitary integration are adapted to solve for the
density matrix as a function of time, including the non-unitary effects
of dissipation and decoherence. The input requires only solutions of
classical, initial value time-dependent equations. 
Results are illustrated for a damped,
driven two-level system.
\end{abstract}

\pacs{03.65.Yz, 05.30.-d, 42.50.Lc}

\maketitle

The study of open quantum systems is of widespread interest across different
areas of physics particularly in the irreversible processes of dissipation
and decoherence afforded by coupling to an external reservoir or
environment. Quantum optics is replete with such studies for optical
bistability, resonance fluorescence, and the general evolution from pure to
mixed states, often considered through damped, driven two-level atoms \cite{ref1}.
Coupled quantum wells in a wider context and the study of quantum Brownian
motion, dissipation and fluctuations have also received much attention \cite{ref2}.
Application of such considerations to ``quantum non-demolition'' in the
emerging field of laser-interferometric gravitational wave detection, and of
quantum noise and decoherence in the field of quantum computation, add to
the importance of this subject. Finally, this evolution from pure to mixed
states is at the heart of the problem of measurement in quantum theory \cite{ref3}.

On the other hand, unitary integration schemes for the evolution operator of
time-dependent Hamiltonians, when available, are powerful because they
preserve invariants and are stable, also in numerical application. In this
Letter, we present a general procedure and illustrate with an example how to
preserve most of these advantages even while working with systems exhibiting
dissipation and decoherence. There are two key steps. First, the $n$%
-dimensional Liouville-von Neumann-Lindblad (LvNL) equation containing
dissipation and decoherence is embedded in a $(n^2-1)$-dimensional
Liouville-Bloch form with a non-Hermitian Hamiltonian. Second, this
Liouville-Bloch equation is handled by a ``unitary integration'' procedure that has been described in recent years \cite{ref4,ref5,ref6} wherein the evolution operator is written as a product of exponentials, each exponent involving an element of a closed Lie algebra of operators together with a multiplicative classical function of time. With all the non-commutativity handled analytically, the entire problem is reduced to solving coupled, first-order differential
equations for this set of \textit{classical }functions. In many cases, this
set reduces to a single non-trivial Riccati (first order, quadratically
nonlinear) equation for one of the classical functions, all the rest then
obtained through trivial quadratures \cite{ref6}. All of the above features remain valid even when the Hamiltonian is non-Hermitian and the evolution non-unitary.

Two other papers share our aims in setting the passage from pure to mixed
states in a unitary evolution scheme but they proceed differently. One deals with
weak dissipation, handling the Hermitian part of the LvNL equation through
unitary integration and the dissipative terms through conventional
integrators \cite{ref7}. Because of their focus on numerical integration, both these
handlings are for small time steps whereas we aim for integration over
arbitrary, finite $t$. Another work \cite{ref8} introduces a novel ``square root
operator'' of the density matrix and an associated $n^2$-dimensional Hilbert
space, along with additional constraints that are not in conventional
quantum mechanics. Our embedding in a higher dimensional space does not
introduce any new elements beyond those already in the density matrix.
After submitting our Letter, we have learnt of another work that solves
master equations by invoking an ``auxiliary" $n^2$-dimensional Hilbert
space \cite{ref9}.

We begin with the master equation for the density matrix $\rho $, sometimes
called the Liouville-von Neumann-Lindblad equation \cite{ref1, ref2, ref3},

\begin{eqnarray}
i\dot{\rho} & = & [H,\rho ]+
\frac{1}{2}i\!\sum_{k}\left( [L_k\rho,L_k^{\dagger }]+
[L_k,\rho L_k^{\dagger }]\right)  \nonumber \\
& = & [H,\rho ]-\frac{1}{2}i\!\sum_{k}\left( L_k^{\dagger }L_k\rho +\rho
L_k^{\dagger }L_k-2L_k\rho L_k^{\dagger }\right)\!, \label{eqn1}
\end{eqnarray}

\noindent where an over-dot denotes differentiation with respect to time and 
$\hbar $ has been set equal to unity, $H$ is a Hermitian Hamiltonian, and
the second term on the right-hand side is the ``Liouvillian super-operator''
describing coupling to the environment and the resulting irreversibilities
of dissipation and decoherence. The above form in the Markov approximation
with an explicitly traceless right-hand side guarantees conservation of $Tr$(%
$\rho $) and positivity of the probabilities. For a more mathematical
description in terms of so-called ``dynamical semigroups,'' we refer to
\cite{ref10,ref11}.

Our aim in this paper is to solve Eq.~(\ref{eqn1})  for fairly general
time-dependences of $H$ and the $L$'s contained in it, while keeping as
closely as possible to the unitary integration that applies in the absence
of the super-operator. This method \cite{ref4,ref5,ref6} has been developed when $H(t)$ is a
sum of terms, each of which involves a time-independent operator multiplying
a classical function of time. In such a case, without any recourse to
time-ordered Dyson expansions, one can solve for the evolution operator $%
U(t) $ satisfying

\begin{equation}
i\dot{U}(t)=H(t)U(t),\quad U(0)=\mathcal{I},  \label{eqn2}
\end{equation}

\noindent by writing $U(t)$ as a product

\begin{equation}
U(t)=\prod_{j}\exp [-i\mu _j(t)A_j],  \label{eqn3}
\end{equation}

\noindent where $A_j$ are the operators contained in $H(t)$ together with a
sequence of other operators formed out of their mutual commutators in a
successive fashion. If this set forms a closed algebra under commutation,
then upon substitution, Eq.~(\ref{eqn3}) can be shown to satisfy Eq.~(\ref{eqn2}) through
repeated application of the Baker-Campbell-Hausdorff (B-C-H) identity [4,6].
This results in a well defined set of coupled first-order, generally
nonlinear, equations for the functions $\mu _j(t).$ Thereby the quantal
problem is reduced to the classical one of solving this set of equations,
following which $\rho (t)$ is obtained as

\begin{equation}
\rho (t)=U(t)\rho (0) U^{\dagger }(t).  \label{eqn4}
\end{equation}

In extending this procedure to non-unitary evolution, if we were to retain only the first two
terms in the superoperator, 
it is simple to extend
Eq.~(\ref{eqn4}) by using two different products $U_L(t)$ and $U_R(t)$ so that $\rho
(t)=U_L(t)\,\rho (0)\,U_R^{\dagger }(t),$ with correspondingly different
functions $\mu _{Lj}(t)$ and $\mu _{Rj}(t)$ in Eq.~(\ref{eqn3}). Once again, upon
calculating $i\dot{\rho}$ with such a form, the B-C-H identity can be used
to 
get a well-defined set of equations for the $\mu _L$ and $\mu
_R.$ However, the last term in the superoperator in Eq.~(\ref{eqn1}), wherein $\rho
(t)$ occurs between operators multiplying it both on the right and from
the left, no longer permits easy generalization. Note that this last term is
the so-called ``quantum jump'' in interpretations of the LvNL equation as
conventional continuous evolutioni, albeit with a non-Hermitian Hamiltonian, 
plus a jump \cite{ref12}.

For the full master equation, we proceed by separating the invariant $%
Tr(\rho )$ from the $n^2$ elements $\rho _{ij}(t)$. Eq.~(\ref{eqn1}) then reduces
for the remaining $n^2-1$ elements to the Liouville-Bloch form

\begin{equation}
i\dot{\eta}(t)=\mathcal{L}(t)\eta (t),  \label{eqn5}
\end{equation}

\noindent where one convenient choice for the $(n^2-1)$ elements of $\eta $
is $\rho _{11}-\rho _{ii}$, $i=2,3,\ldots ,n;$ $\rho _{ij}+\rho _{ji},$ $\rho
_{ij}-\rho _{ji},$ $i>j.$ The first $(n-1)$ of these describe the diagonal
elements of the density matrix, the other $(n^2-n)$ $i\neq j$, describe,
respectively, in-phase dispersive and out-of-phase absorptive components of
polarization. Even though $\mathcal{L}$ may not be Hermitian, the form of
Eq.~(\ref{eqn5}) is now the same as in Eq.~(\ref{eqn2}) with all operators to the left of $%
\eta $ so that the same procedure of a product exponential form for $\eta
(t) $ as in Eq.~(\ref{eqn3}) can be carried out now in the $(n^2-1)$-dimensional
space. Thereby, the LvNL equation for $\rho $ has been embedded in a
higher-dimensional Liouville-Bloch equation. While invariants are no longer
preserved with $\mathcal{L}$ non-Hermitian, the advantages of exponential
factors, with all operator aspects handled analytically and only classical time-dependent equations to solve, still remain.

One immediate consequence is worth noting. If the operators $L_k$ in
Eq.~(\ref{eqn1}) are such that $\mathcal{L}$ in Eq.~(\ref{eqn5}) involves imaginary elements
and, consequently, $\eta $ decays asymptotically, $\eta (t\rightarrow \infty
)\rightarrow 0$, then all coherences vanish (off-diagonal $\rho _{ij}$) and
all diagonal $\rho _{ii}$ become equal, $\rho _{ii}(t\rightarrow \infty
)\rightarrow (1/n)Tr(\rho (0))$. $Tr(\rho ^{2})$ on the other hand, decreases asymptotically to $(1/n)$ of its initial value. A specific $n=2$ illustration will be given below of this rather general conclusion.

To demonstrate this method, we turn now to a series of recent papers \cite{ref13}
that discussed phase coherences and transitions in a periodically driven
two-level system with a single $L$ in Eq.~(\ref{eqn1}):

\begin{equation}
H=\frac{1}{2}\epsilon (t)\sigma _z+J\sigma _x ,\; L=\sqrt{\Gamma }%
\sigma _z ,\; \rho _{ij}(0)=\delta _{ij}\delta _{i1}.  \label{eqn6}
\end{equation}

\noindent Applying our procedure, we have $\rho _{11}(t)+\rho _{22}(t)=1$,
and Eq.~(\ref{eqn5}) for the three remaining elements takes the form

\begin{eqnarray}
i\frac {d}{dt}
\left( 
\begin{array}{c}
\rho _{12}+\rho _{21} \\ 
\rho _{21}-\rho _{12} \\ 
\rho _{11}-\rho _{22}
\end{array}
\right) & = & \left( 
\begin{array}{ccc}
-i\Gamma & -\epsilon (t) & 0 \\ 
-\epsilon (t) & -i\Gamma & 2J \\ 
0 & 2J & 0
\end{array}
\right)  \nonumber \\ 
& &\quad \quad \times \left( 
\begin{array}{c}
\rho _{12}+\rho _{21} \\ 
\rho _{21}-\rho _{12} \\ 
\rho _{11}-\rho _{22}
\end{array}
\right).
\label{eqn7}
\end{eqnarray}

\noindent To solve this as a product of exponentials, we need the eight
operators of an SU(3) algebra. Instead, we illustrate first a simplified variant of
Eq.~(\ref{eqn6}) as our model, with a symmetric choice for the $L_k$ involving all
three Pauli matrices, that is, $L_k=\sqrt{\Gamma /2 }\sigma _k.$ This
modifies Eq.~(\ref{eqn7}) to introduce also a $(-i\Gamma )$ in the third diagonal
element of the matrix. With the matrix then expressible as

\begin{equation}
\mathcal{L}=-i\Gamma \mathcal{I}-\epsilon (t)A_z+2JA_x\;,  \label{eqn8}
\end{equation}

\noindent where $A_x$, $A_y$, $A_z$ are the operators of angular momentum in
a representation

\begin{eqnarray}
A_x=\left( 
\begin{array}{ccc}
0 & 0 & 0 \\ 
0 & 0 & 1 \\ 
0 & 1 & 0
\end{array}
\right)&,& A_y=\left( 
\begin{array}{ccc}
0 & 0 & -i \\ 
0 & 0 & 0 \\ 
i & 0 & 0
\end{array}
\right),\, \nonumber \\
A_z &=& \left( 
\begin{array}{ccc}
0 & 1 & 0 \\ 
1 & 0 & 0 \\ 
0 & 0 & 0
\end{array}
\right), 
\label{eqn9}
\end{eqnarray}

\noindent the closed Lie algebra of these three suffices to solve Eq.~(\ref{eqn5})
by our unitary integration procedure. Since this procedure rests only on the
commutators between $A_j,$ we can use any representation of them as is
convenient. We exploit this in choosing Eq.~(\ref{eqn9}) so that $\mathcal{L}$
involves the $A_j$ only linearly. Although, for comparison with \cite{ref13}, only $%
\epsilon $ in Eq.~(\ref{eqn8}) is a function of time, we note that everything that
follows applies also to more general time dependences of $J$ and $\Gamma $
and inclusion of a time-dependent term in $A_y$ as well. We also note that
reduction of the term involving $\Gamma $ in Eq.~(\ref{eqn8}) to a unit
operator reflects a general sum rule in any dimension $n$. When $k$ in 
Eq.~(\ref{eqn1}) runs over all $n^2$ linearly independent operators, that
summation reduces to $2(n\rho_{ij} -\delta_{ij})$.

The first term in Eq.~(\ref{eqn8}) leads to a trivial factor $\exp (-\Gamma t)$ and
the remaining Hermitian part of $\mathcal{L}$ has been solved before \cite{ref6}:

\begin{eqnarray}
\eta (t)& = &\exp [-\Gamma t] \exp [-i\mu _{+}(t)A_{+}] \nonumber \\
&& \times 
\exp [-i\mu_{-}(t)A_{-}] 
\exp [-i\mu (t)A_z]\eta (0),  \label{eqn10}
\end{eqnarray}

\noindent with $A_{\pm }\equiv A_x\pm iA_y,$ $\eta (0)=(0,0,1),$ and

\begin{subequations}
\begin{eqnarray}
\dot{\mu}_{+}-i\epsilon (t)\mu _{+}-J(1+\mu _{+}^2)
&=&0 ,  \label{subeq1} \\
\dot{\mu}=2iJ\mu _{+}-\epsilon (t) , &&  \label{subeq2} \\
\dot{\mu}_{-}-i\dot{\mu}\mu _{-}=J ,\; && \mu _i(0)=0.  \label{subeq3}
\end{eqnarray}
\end{subequations}

\noindent The first of these equations, involving $\mu _{+}(t)$ alone
in Riccati form, is the only non-trivial member of this set. Solutions give
through Eq.~(\ref{eqn10}),

\begin{eqnarray}
\rho _{11}(t) &=&\frac {1}{2}+\frac {1}{2}\exp (-\Gamma t)\,[1-2\mu _{+}(t)
\mu _{-}(t)],  \nonumber \\
\rho _{22}(t) &=&\frac {1}{2}[1-\exp (-\Gamma t)]+\mu _{+}(t)\mu _{-}(t)
\exp (-\Gamma t),  \nonumber \\
\rho _{12}(t) &=&i\mu _{-}(t)\exp (-\Gamma t),  \nonumber \\
\rho _{21}(t) &=&i\mu _{+}(t)[\mu _{+}(t)\mu _{-}(t)-1]
\exp (-\Gamma t).  \label{eqn12}
\end{eqnarray}

\noindent These are general solutions, valid for any time. The coherences vanish asymptotically and $\rho _{11}$ and $\rho
_{22}$ attain the value $\frac {1}{2}$ as $t\rightarrow \infty $. While $Tr(\rho )$ remains always at unity, $Tr(\rho ^{2})$ decreases to $(1/2)$. The
above assumed as initial state the pure state with $\rho_{11}(0)=1$ the
only non-zero elememt, but a wider choice also leads to the same final
result. Simple
numerical integration of Eq.~(\ref{subeq1}) for an oscillating driving field $%
\epsilon (t)=A\cos (\omega t)$ are shown in Figs.\ 1 and 2 for various values of the parameters $(\omega ,J,A,\Gamma )$ . They are in agreement with \cite{ref13}.
In Fig.\ 2(c), we also record the time evolution of the entropy, $S = - Tr(\rho \, \ln \, \rho )$. The value of $\Gamma$ governs the rate of rise as $S$
increases monotonically from 0 to its asymptotic limit of $\ln 2$.

\begin{figure}
\includegraphics[width=3in]{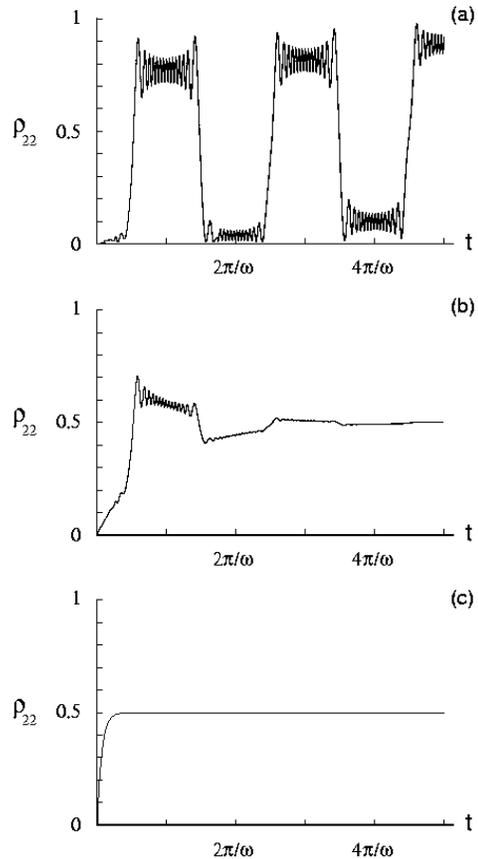}
\caption{$\rho _{22}(t)$ for an oscillating driving field with $J / \omega = 3$, $ A / \omega = 45 $, and damping values (a) $\Gamma / \omega = 0$, (b) $\Gamma / \omega = 0.35$, and (c) $\Gamma / \omega = 5$. }
\end{figure}

We already noted from the $3 \times 3$ matrix structure of Eq.~(\ref{eqn7}) that
for the most general $H$ and $L$ in Eq.~(\ref{eqn6}), a product of eight
exponential operators always provides the requisite $\eta (t)$. As
another illustration of a smaller set sufficing, when only three of the four
linearly independent matrices are included in $L_k$, an additional
inhomogeneous term in the column vector $-i\Gamma (0,0,1)$ appears on the
right-hand side of Eq.~(\ref{eqn5}), with $\mathcal{L}$ again as in
Eq.~(\ref{eqn8}). For $J=0$, this is easily solved, diagonal and off-diagonal
elements decoupling, and gives the result that a mixed state evolves to the 
pure state (1,0).

The reduction in the number of exponential factors required is a generic feature, whenever $\mathcal{L}$ in Eq.~(\ref{eqn5}) involves only the elements of a
sub-algebra of the full algebra of SU($n^2-1$). Thus, in the $n=2$ examples
considered above, the existence of SU(2) subalgebras allows solutions with
just three exponential operators in Eq.~(\ref{eqn10}). Denoting the eight
operators of SU(3) by $O_i, i=1-8$, with one choice for them being
($A_z, A_{+}, A_{-}, A_3 ^2, A_{+} ^2, A_{-} ^2, A_{+}A_3 +A_3 A_{+},
A_{-}A_3 +A_3A_{-}$), there are several triplets that close under commutation.
These include the familiar $i=(1,2,3)$ as in Eq.~(\ref{eqn8}) but also many
others such as (1,5,6) and (1,7,8). There are also sub-algebras involving four
(for example, (1,2,5,7) and (1,3,6,8)) and five elements (examples: (1,2,4,5,7)
and (1,3,4,6,8)) in which case four or five exponential factors, respectively,
would suffice for our solution in Eq.~(\ref{eqn10}). As $n$ increases,
although the total number of operators $n^2-1$ grows rapidly, once again,
$\mathcal{L}$ may involve only the operators of sub-algebras, SU($n^2-1$)
containing many sub-algebras of lower order all the way down to SU(2) with
just three operators. Indeed, with increasing $n$, there are many more such
sub-algebras so that very often the $H$ and $L_k$ may afford reduction of
the number of exponentials in our procedure to a small number.

In summary, an $n$-dimensional LvNL equation describing dissipation and
decoherence (or, alternatively, continuous evolution plus a quantum jump) of
the density matrix $\rho (t)$ is first embedded into an ($n^2-1$)-dimensional
Liouville-Bloch equation for diagonal and off-diagonal combinations
$\eta (t)$ of $\rho (t)$. A unitary integration scheme is then
applied to this form of the equation, with $\eta (t)$ expressed as a product
of exponentials involving a limited, finite number of factors and operators,
often just the three of angular momentum. Through this procedure, all
elements of $\rho (t)$ are obtained in terms of solution of a single Riccati
equation for a classical function together with ordinary multiplication and
integration.

We thank Drs. Dana Browne and Lai Him Chan for suggesting we follow the entropy of evolution.

\begin{figure}
\includegraphics[width=3in]{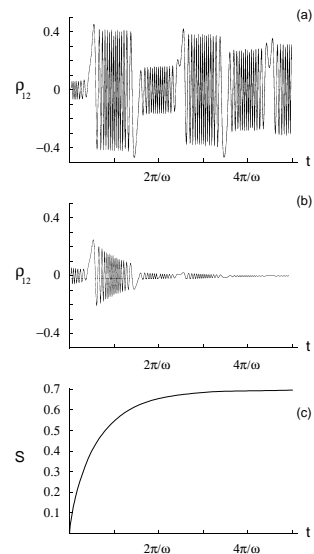}
\caption{As in Fig.\ 1, for $\rho _{12}(t)$ with (a) $\Gamma / \omega = 0$ and (b) $\Gamma / \omega = 0.35$. The entropy $S$ for $\Gamma / \omega = 0.29$ is shown in (c).}
\end{figure}

\end{document}